Peer-reviewed and presented at 2025 International Conference on Advancement in Communication and Computing Technology (INOACC-2025); self-published by the author due to a sustained 13-month indexing delay by the organizers.

# An Efficient Machine Learning-based Framework for Detection and Prevention of Frauds in Telecom Networks

Praveen Hegde
*Associate Director-Emerging Commercial Platforms*
*Verizon*
Atlanta, USA

Mishal Shah
*Senior Software Engineer*
*Bloomberg LP*
Jersey City, USA

***Abstract*—Telecommunication fraud is an acute problem that leads to substantial material losses and compromises the reliability of telecom systems worldwide. Only effective and efficient detection mechanisms can help to deal with these threats, though there are certain shifts in the approaches to fraud detection. This paper evaluates the performance of AI-driven models for fraud detection in telecommunication networks using Call Detail Record (CDR) datasets. This study focuses on fraud detection in telecom networks using the Telecom CDR dataset, which contains 101,174 customer records with 17 attributes, including 8,830 fraud cases. In feature preprocessing, missing values were dealt with, followed by data scaling using Min-Max scaling and data balancing using the SMOTE technique. The dataset was trained for predictive analysis using Random Forest (RF) and XGBoost models. F1-score, ROC AUC, recall, accuracy, time, and precision were used as indicators with which to compare performance of the two models. RF recorded a high level of accuracy at 99.9% while XGBoost at 99.7%. Findings show that the suggested framework successfully detects fraud with few misclassifications. Several machine learning models were evaluated and contrasted, such as RF, XGBoost, DBSCAN, RoBERTa, and K-means. Among all the models, RF was seen to give the highest performance with an accuracy of 99.9% and precision of 99.9%, recall of 99.9% and F1-score of 99.9%, XGBoost, GNN and BERT. The findings emphasize RF as the most effective model for detecting fraudulent activities in telecom networks, ensuring robust and reliable prevention of fraud.**

## I. INTRODUCTION

Online banking, e-commerce credit card transactions, and telecommunication fraud are just a few examples of the many fraudulent activities that have flourished in recent years, thanks to the proliferation of new technology. These scams cost businesses and consumers across the globe over a billion dollars annually [1]. In truth, fraud is quite expensive for all telecom operators when it comes to capacity and revenue lost [2]. A study by Neural Technologies estimated that in 2016, the telecom sector incurred an average loss of $249 billion due to fraud [3]. Fraud may be described as the unlawful utilization of telecommunications infrastructure, such as mobile communications, with the goal of not paying for services, abusing voice calls (or data, SMS, or MMS), deceiving subscribers, and unlawfully accessing telecommunications provider networks[4][5]. The convergence of telecommunications and financial services has created a dynamic landscape where mobile and digital transactions are increasingly prevalent[6]. However, this growth is not without issues on cybersecurity risks, particularly in the aspect of fraud prevention [7][8]. Since financial transactions have become closely connected with telecommunication networks, instances of fraud activity have increased[9], thus requiring the use of increased security for protecting significant financial data and ensuring the purity of financial systems[10].

Telecommunication network fraud may present itself in several forms such as breaking into the devices that form part of the network,[11] interfering with the routing protocols, exploiting the weakness in the networking and so on[12]. These activities not only incur tangible financial but also lead to erosion of confidence and certainty in the telecommunication services [13]. Thus, telecom fraud can be performed in a wide range of ways with reference to the capabilities and opportunities offered to the fraudster [14][15]. A wide number of techniques is used for fraud detection. Fortunately, Artificial intelligence based Machine Learning (ML) technologies have become an effective weapon in the fight against telecom fraud, nowadays[16][17]. ML techniques have made it possible to develop smart methodology using data, models for learning from data and making predictions for automated inspection-free processes [18][19]. ML algorithms have the ability to analyze enormous amounts of data, and using ML-based classification allows to assess malicious actions and protect against fraud.

### *A. Motivation and Contribution of the Study*

Telecom fraud poses significant financial and operational challenges, necessitating advanced solutions to identify and mitigate fraudulent activities effectively. Traditional methods often struggle with scalability and accuracy in handling large, imbalanced datasets. This research introduces a novel ML framework that combines robust preprocessing techniques, such as SMOTE for data balancing, with powerful predictive models like Random Forest and XGBoost. By benchmarking against emerging methods like BERT and GNNs, the framework not only enhances detection capabilities but also provides insights into model suitability, addressing gaps in existing research. The aim is to develop an efficient and reliable system for detecting and preventing telecom fraud, offering a significant improvement over existing approaches, through comprehensive evaluation and benchmarking. The following are the primary benefits of this research:

- The study leverages a substantial and realistic Telecom CDR dataset with diverse attributes, providing a strong

basis for developing fraud detection models tailored to the telecom industry.

- The adoption of preprocessing techniques, including missing value handling, Min-Max scaling, and SMOTE-based data balancing, ensures a robust and unbiased dataset for model training and evaluation.
- By implementing and analyzing multiple ML approaches, including RF and XG Boost, the study contributes to understanding their relative suitability for telecom fraud detection.
- To evaluate the efficacy of the model in identifying telecom fraud, the research used a comprehensive set of performance indicators, including accuracy, precision, recall, F1-score, and ROC AUC.
- The proposed methodology demonstrates scalability and adaptability, making it a viable solution for deploying fraud detection systems in real-world telecom networks.
- The research compares its models to modern methods like DBSCAN, RoBERTa, and K-means, shedding light on their relative merits and shortcomings when it comes to detecting fraud.

### B. *Organization of the paper*

Here is the structure of the paper: The literature study on telecom fraud and the obstacles in detecting it is covered in Section II. The methods and models used in the research are detailed in Section III. The experimental results and performance analysis are detailed in Section IV. The paper concludes in Section V, which explores possible future study directions.

## II. LITERATURE REVIEW

This section of the research summarises previous efforts that have addressed the topic of telecom network fraud categorization and detection. The majority of the publications under consideration concentrate on machine learning methods and frameworks that have been created to better identify and stop telecom fraud. Some reviews are:

H. palivela et al. (2024), research aims to identify transaction patterns to help algorithms recognize and flag scammers. The proposed approach incorporates methods of ensemble learning, including Gradient Boost, Random Forest, Logistic Regression, and Voting Classifiers, along with hyperparameter tuning to prevent overfitting. It effectively manages false positives through careful review, achieving an accuracy of 99.59%, surpassing other innovative models. [20].

Jun Li et al. (2024) proposed RoBERTa-MHARC, a model for text-based telecom fraud detection that combines RoBERTa with residual connections and a multi-head attention feature. The approach begins by sorting the CCL2023telecom fraud dataset into many groups. Experimental results demonstrated that our model attains better results than the benchmarks on several datasets include an F1 score97.65 on the FBS dataset, 98.10 on our own dataset, and 93.69 on the news dataset [21].

V. Chang et al. (2024), this research looked at how well the SMOTE performed when compared to two other methods: random under-sampling and over-sampling. The results show that random under-sampling achieves good recall (92.86%) but poor precision, whereas SMOTE makes a better accuracy (86.75%) and more consistent F1 score (73.47%), but its recall is somewhat lower[22].

R. Li et al. (2023), provide a method for reducing attributes that takes into account MCIR in order to enhance the quality of the data. They proceed to address partial or missing data by developing a MCIR-RGAD based on the concept of maximum consistent blocks. Finally, the MCIR-RGAD method is shown to be an efficient solution for lowering computing time, enhancing data quality, and processing incomplete data in experimental findings using actual telephony fraud data and UCI data[23].

S. K. Hashem et al. (2022), to manage the relative importance of valid and fraudulent transactions, you may want to think about class weight-tuning hyperparameters. The findings display that LGBM and XGBoost reach the best level requirements with ROC-AUC=0.95, precision0.79, recall0.80, F1 score0.79, and MCC0.79. Through the use of DL and the Bayesian optimization technique, we are also able to achieve the following criteria: ROC-AUC=0.94, precision=0.80, recall=0.82, F1 score=0.81, and MCC=0.81[24].

Table 1 presents the comparative analysis of linked studies according to their Key Findings, limitations, future work, and Focus On.

TABLE I. SUMMARY OF RELATED WORK FOR TELCO FRAUD DETECTION USING MACHINE LEARNING TECHNIQUES

| Reference | Methodology | Dataset | Results | Limitations | Future Work |
|---|---|---|---|---|---|
| H. Palivela et al. (2024) | Ensemble learning methods: Gradient Boost, Random Forest, Logistic Regression, and Voting Classifiers with hyperparameter tuning. | Sizable dataset of anonymized credit card transactions | Achieved an accuracy of 99.59%, surpassing other models. | Cannot entirely prevent fraudulent transactions; relies on post-detection manual review. | Enhance scalability and real-time detection to further reduce false positives. |
| Jun Li et al. (2024) | The RoBERTa-MHARC paradigm encompasses RoBERTa, a multi-head attention mechanism, and residual connections. | CCL2023 telecom fraud dataset and benchmark datasets | F1 score of 97.65 (FBS dataset), 98.10 (own dataset), 93.69 (news dataset). | Requires robust preprocessing for varied dataset quality; computationally intensive. | Extend the model to real-time and multilingual fraud detection scenarios. |
| V. Chang et al. (2024) | Comparison of random under-sampling and SMOTE for class balancing in fraud detection models. | Fraudulent transaction datasets. | Random under-sampling: recall of 92.86%. SMOTE: accuracy of 86.75%, F1 score of 73.47%. | Random under-sampling compromises precision; SMOTE introduces synthetic data, which may not reflect real-world cases. | Investigate advanced balancing techniques and their integration with deep learning models. |
| R. Li et al. (2023) | Attribute reduction algorithm (MCIR) and rough-gain anomaly detection (MCIR-RGAD) for handling incomplete data and improving data quality. | Telecommunication fraud data and UCI dataset. | Reduced computation time, improved data quality, and effective processing of incomplete data. | Limited scalability to large datasets; focused mainly on incomplete data scenarios. | Extend to more diverse datasets and improve scalability for large-scale implementations. |

| S. K. Hashem et al. (2022) | Class weight-tuning hyperparameters, majority voting ensemble learning method, LightGBM, XGBoost, and Bayesian optimization for hyperparameter tuning. | Fraudulent transaction datasets. | LightGBM and XGBoost: ROC-AUC = 0.95, precision = 0.79, recall = 0.80, F1 score = 0.79, MCC = 0.79. | Requires extensive computational resources; ensemble methods may increase complexity and reduce interpretability. | Apply these methods to real-time fraud detection systems and reduce computational overhead. |
|---|---|---|---|---|---|

## III. METHODOLOGY

This research aims to assess models powered by artificial intelligence for the purpose of detecting fraud in telecommunications networks. The flow diagram in Figure 1 shows the research methodology. The process begins with the collection of a telecom Call Detail Record (CDR) dataset containing key attributes such as account length, call minutes, charges, and fraud status. Initially, the dataset underwent preprocessing, including handling missing values, noise reduction, and balancing data using the SMOTE technique. Following this, feature scaling was performed using Min-Max normalization to the dataset. The data that had already been preprocessed was then divided into two parts: training and testing. At last, we put a number of ML models into action, including XGBoost and RF, and we measured their efficacy using measures including AUC, F1-score, recall, precision, and accuracy. The outcomes provide a comprehensive comparison of these models with DBSCAN, RoBERTa and K-means in detecting fraudulent activities effectively.

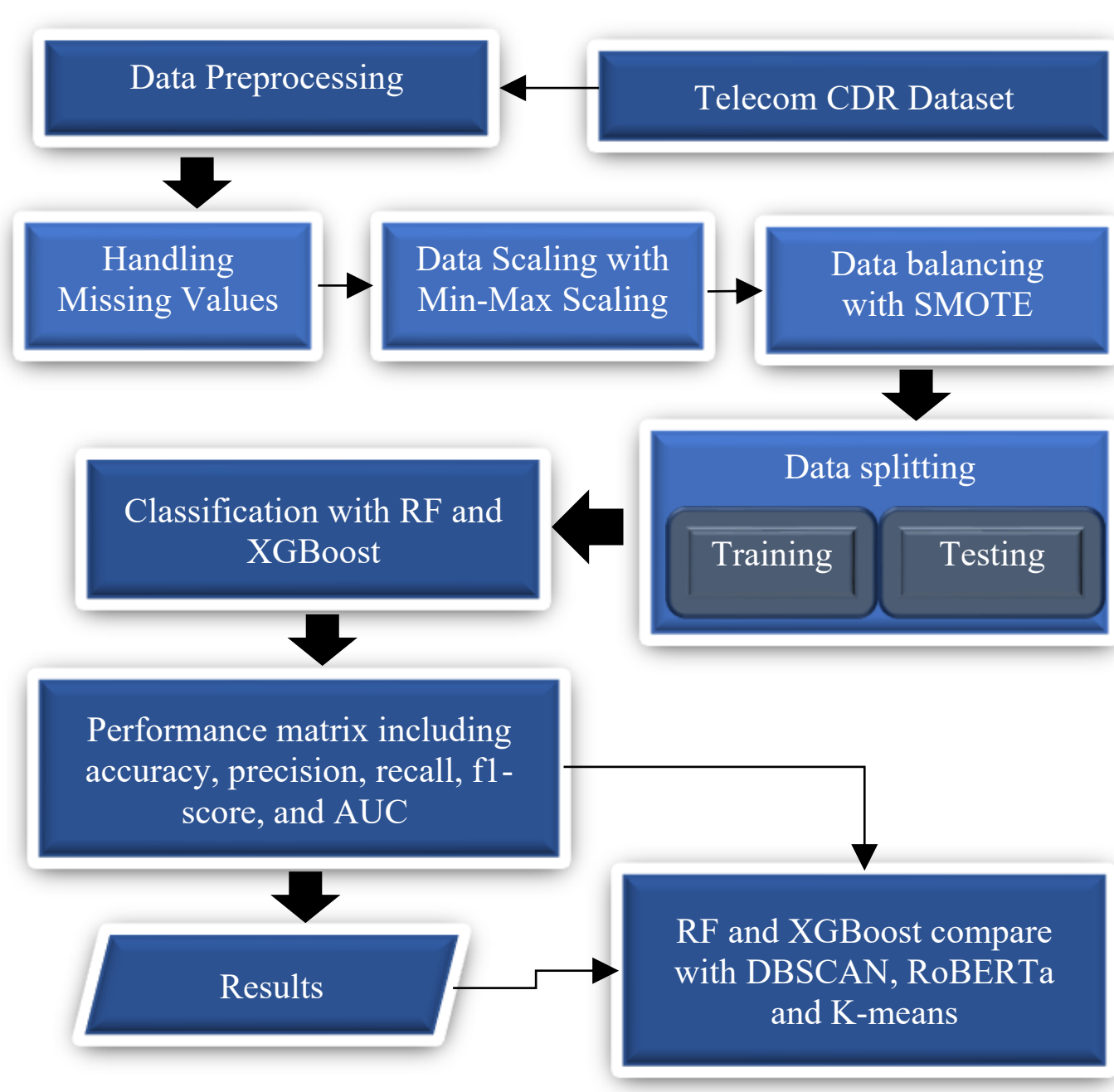

Fig. 1. Flowchart for Detection of Frauds in Telecom

The overall steps of the flowchart for of machine learning models for fraud detection are provided in below:

### A. Data Collection

This study used the Telecom CDR Dataset[1]. The dataset used for this study includes 17 attributes from 101,174 customers, with 8,830 fraud, shows in figure 2. The following variables are included in it: state, account duration, phone number, international plan, mail plan, amount of voice mail messages, call minutes, costs for day, evening, and night, international use, total income from SMS, and current status of fraud.

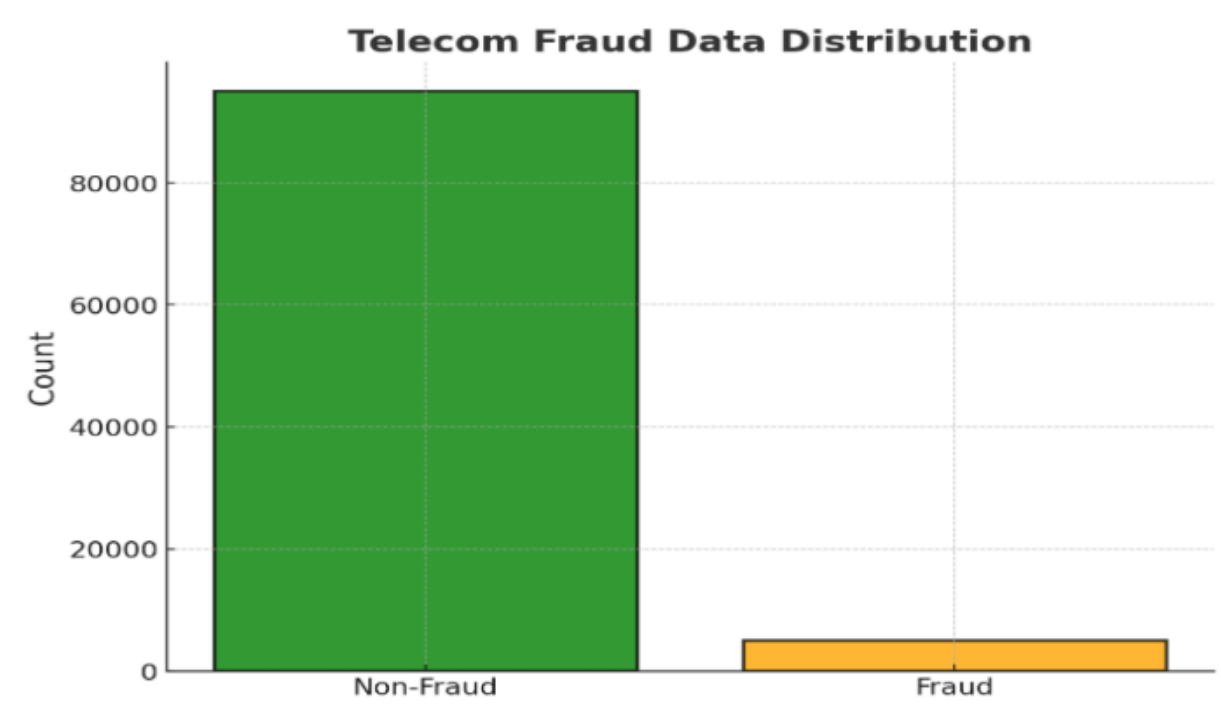

Fig. 2. Telecom Fraud Data Distribution

### B. Data Preprocessing

The concept of ''data preprocessing'' is used to describe any action taken on unprocessed data before it is used. Data preparation is the process of transforming raw data into a more usable format [25][26]. Raw data undergoes a series of treatments known as data preparation to make it more useable and processed. Below is a list of the pre-processing steps:

- Handling Missing Values: Missing values were identified and removed to maintain the integrity of the dataset. Transactions with missing critical features were excluded.
- Data Scaling with Min-Max Scaling: is that the features will be rescaled to ensure the[27] mean and the standard deviation to be 0 and 1, respectively. The Min-Max Scaling formulate as equ.1.

$$x' = x - \frac{(x_{min})}{(x_{max} - x_{min})} \tag{1}$$

where X is an initial value, X' is a transformed value, and Xmax and Xmin are a highest and lowest points for that characteristic, respectively.

### C. Data balancing with SMOTE

The models were made more accurate by using the SMOTE. To improve the representation of under-represented groups in datasets, SMOTE employs an oversampling technique that, given an existing class sample, creates additional samples from that sample [28]. The technique generates non-duplicative minority class samples by convexly combining two or more neighboring data samples in the feature space that are chosen at random. Figure 3 shows the after SMOTE techniques applied balanced classes non-fraud and fraud.

[1] https://github.com/jamesrawlins1000/Telecom-CDR-Dataset-

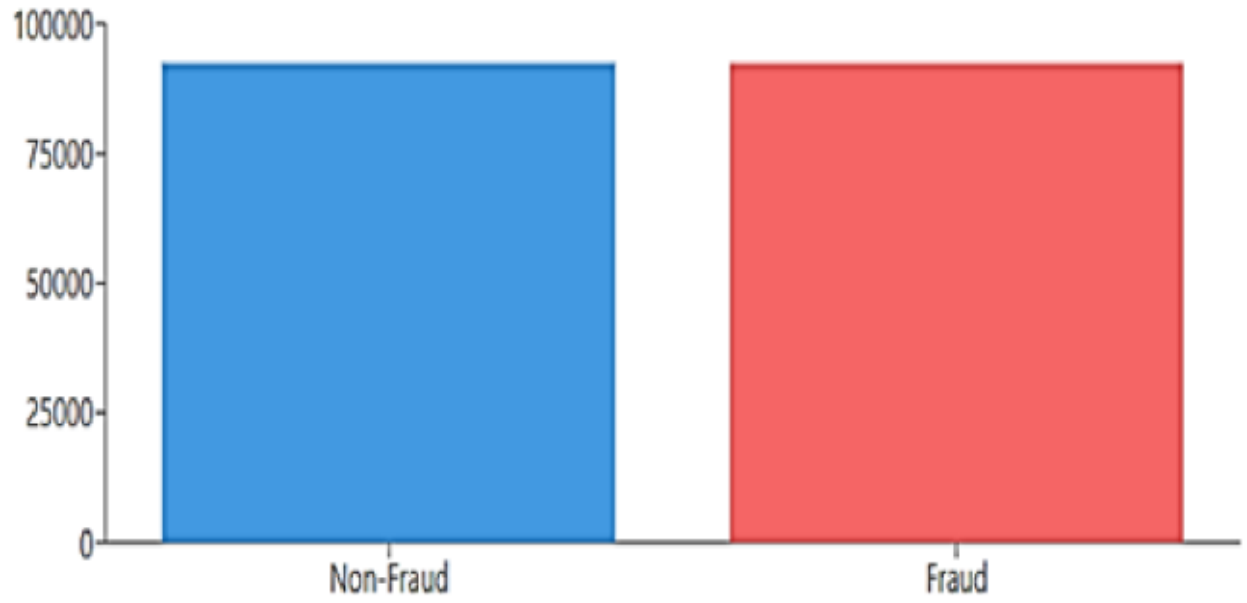

Fig. 3. Telecom Fraud Data Distribution

### *D. Data splitting*

Data splitting involves separating a dataset into smaller parts so that each phase may be trained, validated, and tested independently. The training set is used to train the model, while the test set is used to evaluate its performance. Two portions of the dataset were separated: 80% for training and 20% for testing.

### *E. Prediction with ML-based RF and XGBoost model*

In this study, have utilized both RF and XGBoost models for predictive analysis. RF leverages multiple DT to provide robust predictions for regression and classification, optimizing loss functions like squared error and zero-one loss. XGBoost, known for its efficiency, enhances accuracy through iterative loss and regularization minimization. Together, these models ensure high-performance predictions across the dataset.

#### *1) Random Forest*

Trees are the backbone of the RF ensemble, and each tree's performance is affected by a collection of random variables [29]. To put it another way, we may think of the input or predictor variables X= (X1,..., Xp)T as having actual values and the response variable Y as having real values. Then, we can imagine an unknown joint distribution PXY (X, Y) for these variables. The aim is to find a function (X) that can predict Y [30] The prediction function is chosen by the loss function L(Y, f(X)) with the aim of minimizing the anticipated value of the loss equation 2.

$$E_{XY}(L(Y, f(X))) \tag{2}$$

If X and Y are joint distributions and the subscripts indicate expectations with regard to that distribution.

It seems to reason that L(Y,f(X)) would be a measure of the proximity of f(X) to Y; it would penalise f(X)values that are far by Y. Options for Lare's squared error loss In regression, L(Y,f(X)) = (Y−f(X))2 and in classification, zero-one loss is equal to equation 3.

$$L(Y, f(X)) = I(Y \neq f(X)) = \begin{cases} 0 \; if \; Y \; = f(X) \\ 1 \; otherwise \end{cases} \tag{3}$$

Equation 4 is a conditional expectation obtained by minimising EXY (L(Y,f(X))) with respect to squared error loss.

$$f(x) = E(Y|X = x) \tag{4}$$

sometimes called a regression function. Assuming that Y stands for the set of possible Yi values in a classification situation, the Bayes rule equ.5 may be written as the minimum of EXY(L(Y,f(X))) for a loss of zero.

$$f(x) = argmaxP(Y = y|X = x) \tag{5}$$

#### *2) Extreme gradient boosting (Xgboost)*

GBM are among the most effective algorithms in supervised learning, and Xgboost is an implementation of this method [31][32]. This tool is versatile enough to tackle both classification and regression issues. Because of its fast execution speed out of core computing, Xgboost is favored among data scientists [33] The Xgboost operates in the following manner: Consider the following scenario: we have a datasetDS with m features and n examplesDS={($X_i$, $Y_i$ ): i=1……n, $\in \mathbb{R}^m, y_i \in \mathbb{R}\}$. An ensemble tree model may be constructed using equation 6 to forecast the output, which is denoted as yi:

$$Å. i = \emptyset(x_i) = \sum_{k=1}^{K}(x_i), f_k \in \mathfrak{F} \tag{6}$$

The number of trees in the model is represented by K. Finding the optimal collection of functions by minimising the loss and regularisation goal in accordance with equation 7 is necessary to answer the aforementioned equation, where fk denotes the (k-th tree).

$$\mathcal{L}(\emptyset) = \sum_i 1(y_i, Å. \iota) + \dot{\sum}_k \Omega(f_k) \tag{7}$$

The loss function, denoted as l, is a difference between an actual output yi and a projected output y^i. this helps prevent the model from being over-fit, where O is a measure of the model's complexity. together with the following equations 8:

$$\Omega(f_k) = \gamma T + \frac{1}{2}\lambda \parallel w \parallel^2 \tag{8}$$

In equation 8, T stands for the total number of tree leaves, and w for the weight of every leaf.

### *F. Performance Metrics*

These performance measures are used to assess the efficacy and precision of the suggested framework for telecom network fraud detection and prevention based on machine learning, with the use of the Telecom-CDR dataset:

- **TP (True Positive):** Positive sample count that was accurately determined.
- **TN (True Negative):** The amount of negative samples that were accurately labelled as negative.
- **FP (False Positive):** The sum of all samples that were labelled as negative but were really positive.
- **FN (False Negative):** Amount of positive samples that were mistakenly labelled as negative.

1) **Accuracy:** Accuracy is a ratio of a number of correctly classified samples to a total number of samples. The corresponding equation (9) is given below:

$$\text{Accuracy} = \frac{(\text{TP})+(\text{TN})}{(\text{TP + TN + FP + FN})} \tag{9}$$

2) **Precision:** Accurate positive predictions inside a positive class are measured as precision. Precision is best understood as follows, according to mathematical equation (10):

$$Precision = \frac{(TP)}{(TP)+(FP)} \tag{10}$$

3) **Recall:** The imbalance issue relies heavily on the recall metric, which reliably measures the percentage of significant categories that are discovered[34]. Below are the matching equations (11):

$$Recall = \frac{(TP)}{(TP)+(Fn)} \tag{11}$$

4) **F1-Score:** F1 is another all-encompassing indicator for assessing unbalanced issues; it is described as equation (12):

$$F1 - Score = 2 \times \frac{Precision \times Recall}{Precision + Recall} \quad (12)$$

5) **Receiver operating characteristics curve (ROC):** A method for representing the efficacy of a model graphically is the ROC. This index fully captures the sensitivity and specificity factors, which are continuous. The correlation among recall and precision is seen by the curve.

## IV. RESULT & DISCUSSION

The results of the study's classification techniques are examined here, with a particular emphasis on the RF and XGBoost models' capabilities for detecting telecom fraud. These models were benchmarked and validated on the Telecom CDR dataset using precision, accuracy, recall, F1-score and ROC AUC. A Python-based simulation environment using Jupyter Notebook running on Google Colab was used to carry out the experimental results. In the analysis, the required libraries where utilized and these included; Scikit-learn, NumPy, Pandas, Keras, Matplotlib. A local machine with an Intel i7 processor, 16 GB of RAM, and an NVIDIA GTX 1660 Ti GPU for best performance was used to run the simulation. The table II provide the performance of parameters for telco fraud detection.

TABLE II. RF AND XGBOOST MODEL PERFORMANCE ON THE TELECOM CDR DATASET

| Matrix | RF | XGBoost |
|---|---|---|
| Accuracy | 99.9 | 99.7 |
| precision | 99.9 | 99.8 |
| Recall | 99.9 | 99.8 |
| F1-Score | 99.9 | 99.8 |
| ROC AUC | 99.9 | 99.8 |

The table II shows the performance of RF and XGBoost model on the Telecom CDR Dataset. The models achieved 99.9% and 99.7% accuracy respectively and other measures performance for fraud detection.

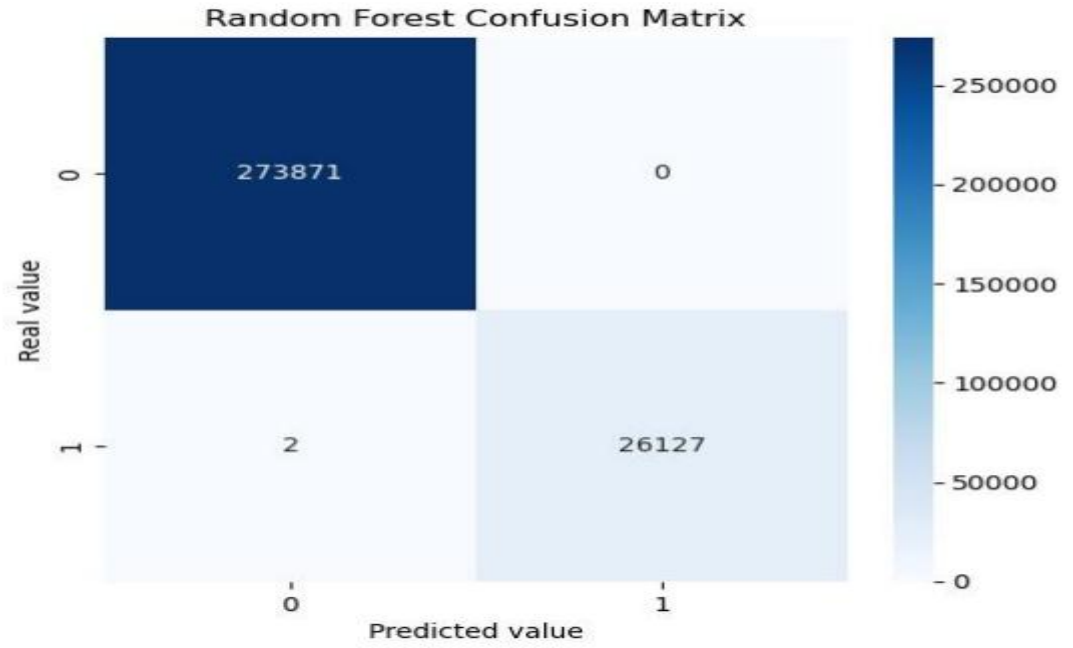

Fig. 4. Random Forest Confusion Matrix

Figure 4 illustrates the confusion matrix, which indicates the Random Forest Classifier's performance. It shows the TP (26127) and TN (273871), indicating the correct predictions for classes 1 and 0, respectively. There are no false positives (0), and only 2 false negatives, showcasing excellent predictive performance.

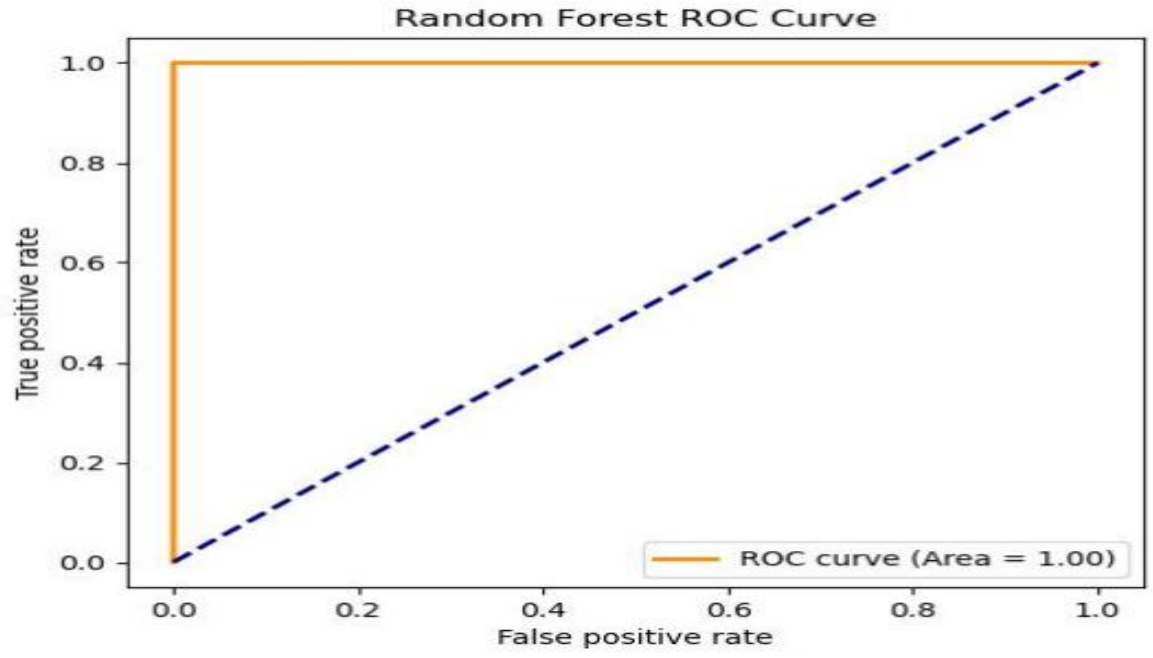

Fig. 5. Random Forest ROC Curve

Figure 5 displays the ROCcurve, which demonstrates how well the Random Forest Classifier differentiates among the classes. The curve closely follows the top-left corner, and the AUC is 1.0, indicating perfect classification.

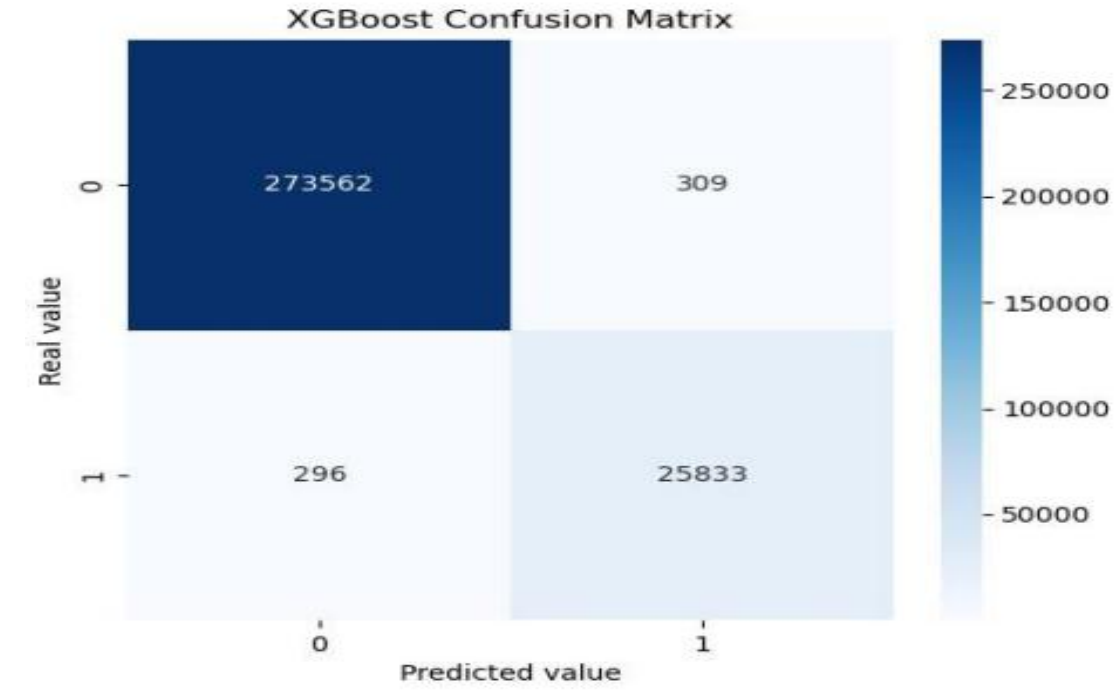

Fig. 6. Confusion Matrix for XGBoost

In figure 6, the confusion matrix represents the performance of an XGBoost classifier. It shows 273,562 true negatives and 25,833 true positives, indicating correct predictions for classes 0 and 1, respectively. There are 309 FP and 296 FN, which are misclassifications. The model works as intended, however there is room for improvement when compared to the ideal categorisation situation due to the occurrence of false positives and false negatives.

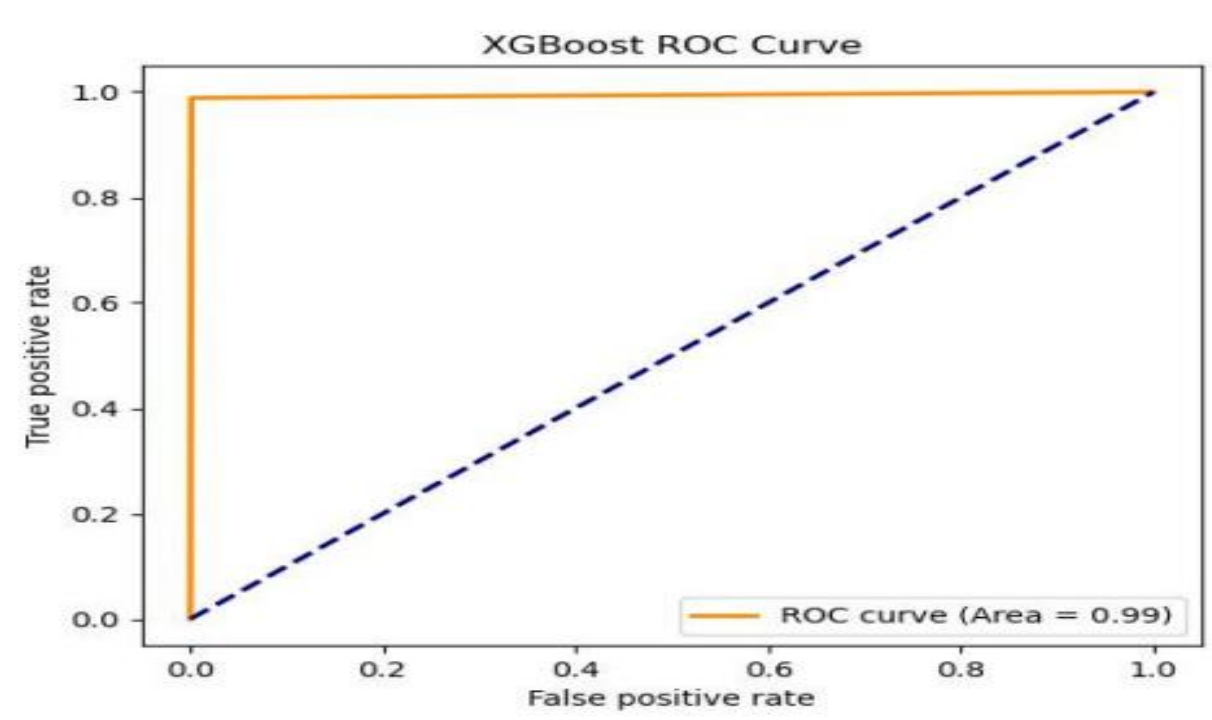

Fig. 7. ROC Curve for XGBoost

Figure 7 shows the ROC curve, which shows how well the XGBoost model performed. The dashed diagonal line stands for random guessing, whereas the orange curve towards the top-left corner signifies good classification performance. The model's exceptional capability to differentiate among positive and negative classes is shown by an AUC of 0.99.

TABLE III. PERFORMANCE COMPARISON OF MACHINE LEARNING MODELS FOR FRAUD DETECTION

| Matrix | DBSCAN [35] | RoBERTa | K-means[36] | RF | XGBoost |
|---|---|---|---|---|---|
| Accuracy | - | 98.10 | 0.961 | **99.9** | **99.7** |
| precision | 0.9039 | 98.11 | 0.471 | **99.9** | **99.8** |
| Recall | 0.9938 | 98.12 | 0.727 | **99.9** | **99.8** |
| F1-score | 0.8989 | 98.12 | 0.533 | **99.9** | **99.8** |

The table below compares the efficiency of several ML models trained on a dataset consisting of call detail recordings in order to identify instances of fraud in telecom networks. XGBoost achieved exceptional performance with an accuracy99.9%, precision99.9%, recall99.9%, and an F1-score99.9%, making it the most effective model for this task. RF followed closely with an accuracy99.7%, precision99.8%, recall99.8%, and an F1-score99.8%, showcasing its robustness. RoBERTa demonstrated high accuracy at 98.10%, precision at 98.11%, recall at 98.12%, and an F1-score of 98.12%, performing well but slightly below tree-based models. K-means clustering, while effective in some scenarios, struggled with a lower accuracy0.961, precision0.471, recall0.727, and an F1-score0.533, highlighting its limitations in detecting fraud. DBSCAN, another clustering technique, showed a recall of 0.9938 and precision of 0.9039 but suffered from a lower F1-score of 0.8989, indicating challenges in handling imbalanced datasets. Overall, XGBoost and RF emerged as the most efficient and reliable models for fraud detection in telecom networks.

## V. Conclusion & Future Work

Fraud detection in telecommunication networks is an ongoing challenge that requires innovative approaches to combat the increasing sophistication of fraudulent activities. The proposed machine learning-based framework for telecom fraud detection shows exceptional predictive performance. The Random Forest classifier achieved near-perfect results with an AUC of 1.0, while XGBoost delivered comparable performance with an AUC of 0.99. This study affirms the effectiveness of ensemble learning approaches, including RF and XGBoost, in managing class imbalance and producing optimal accuracy in detecting fraud ready for further large-scale application in the telecoms networks. As the research shows the feasibility of using Random Forest and XGBoost models with good performance metrics for the detection of telecom frauds, the study has some drawbacks, especially due to the relative imbalance of the given dataset, which often necessitates oversampling tools like SMOTE, which may introduce synthetic biases. Further, due to the dependence on computational resources the models might possess limited supply in larger real-world applications with larger datasets. Future research could consider deep learning architectures, advanced ensemble techniques, and real-time fraud detection systems. To improve interpretability and model flexibility for a wider range of deployment, more advanced data set holding and explainable AI could be implemented.